**Title**: Biphasic Meniscus Coating for Scalable and Material Efficient Quantum Dot Films

**Author List**: Shlok Joseph Paul,[1] Letian Li,[1] Zheng Li,[1] Andrew Kim,[1] Mia Klopfestein,[2] Stephanie S. Lee,[2] Ayaskanta Sahu*[1]

**Affiliations**:
[1] Department of Chemical and Biomolecular Engineering, New York University Tandon School of Engineering, 6 Metrotech Center, Brooklyn, NY, 11201, USA
[2] Molecular Design Institute, Department of Chemistry, New York University, New York, NY, 10012, USA

* Please forward all correspondence to asahu@nyu.edu

**Abstract**:
Colloidal quantum dots (cQDs) have emerged as a cornerstone of next-generation optoelectronics, offering unparalleled spectral tunability and solution-processability. However, the transition from laboratory-scale devices to sustainable industrial manufacturing is fundamentally hindered by spin-coating workflows, which are intrinsically wasteful and restricted to planar geometries. These limitations are particularly acute for high-performance cQDs containing regulated elements such as lead, cadmium, or mercury, where poor material utilization exacerbates both environmental burden and cost. Here we report a biphasic dip-coating strategy that redefines the material efficiency of nanocrystal thin-film fabrication. By utilizing an immiscible underlayer to displace ~88% of the active reservoir volume, we demonstrate a deposition geometry that decouples material consumption from total precursor volume. Infrared PbS photodetectors fabricated via this approach maintain their performance against spin-coated benchmarks while reducing ink consumption by up to 20-fold. Our technoeconomic analysis reveals that this biphasic architecture achieves cost parity at film thicknesses an order of magnitude lower than conventional monophasic dip-coating. Our results establish a low-waste framework for solution-processed materials, providing a viable pathway for the resource-efficient manufacturing of optoelectronic devices.

**Introduction**:

Solution-processed thin films underpin optoelectronics, from optical interference coatings and antireflection layers to sensors, photodetectors, and photovoltaics.[1-3] Colloidal quantum dot (cQD) photodiodes, phototransistors, and solar cells are typically assembled through iterative spin-coating cycles, with each deposited layer driven through a solid-state ligand exchange that replaces long insulating ligands with short-chain species to compact the film and strengthen electronic coupling.[4–6] Repeating this process yields dense films of controlled thickness, but it discards most of the deposited solution, often more than 95% per layer, and becomes increasingly inefficient at larger substrate areas.[7] These issues are compounded by the widespread use of cadmium, mercury, and lead quantum dots, all of which fall outside the Restriction of Hazardous Substances (RoHS) directive.[8] This dependence raises broader concerns about the long-term sustainability of these technologies, particularly if process waste is not rigorously controlled.[9–12] Among scalable routes for forming uniform colloidal films, dip coating offers a particularly simple and material-efficient approach.[13–17] The technique meters a film from a reservoir through a controlled meniscus, avoiding the substantial solution loss inherent to centrifugal thinning in spin coating.[9] Additionally, dip coating is compatible with a wide range of solvent systems, including low-volatility formulations, and readily enables multilayer or graded structures through adjustments to withdrawal speed, dwell time, or reservoir composition.[13,18,19] These features are especially valuable for cQD-based optoelectronic devices, whose performance is extremely sensitive to minor changes in surface chemistry and the packing of the quantum dot film.[20,21]

Paradoxically, laboratory dip coating systems that would let researchers exploit the benefits of dip coating remain prohibitively expensive. Commercial instruments with programmable, multi-step recipes routinely cost several thousand dollars while offering limited flexibility for custom workflows.[22] Recent advances in low-cost motors, open-source microcontrollers, and 3-D printing have enabled do-it-yourself scientific hardware platforms that rival commercial performance at a fraction of the price.[19,22,23] Harnessing these developments, we report an open-source, multistage dip-coater that provides fully programmable workflows for cQD deposition methods employing up to three solvents at a nominal cost of ~$300. As a proof-of-concept, we fabricate dip-coated PbS cQD infrared photodetectors that perform comparably to spin-coated devices.

Despite the afore-mentioned advantages, conventional dip coating faces a fundamental limitation since the reservoir must be filled with a large volume of colloidal dispersion, an increasingly impractical requirement when solvents are unstable or hazardous, or when quantum dot synthesis cannot be readily scaled.[14] These constraints often push researchers toward spin coating, even though it discards most of the active solution. A promising alternative is a biphasic dip-coating strategy in which only a small fraction of the active solution, typically ~10% of the reservoir volume, is floated atop an immiscible underlayer such as perfluorohexane. Originally demonstrated by Grosso *et al.* for single dip sol–gel oxides and block copolymers, this configuration preserved film quality while sharply reducing material use.[14] The approach remains largely untapped for cQDs, despite its clear relevance to a wide range of optoelectronic devices.[13] Here we implement biphasic dip coating for PbS cQDs in infrared photodetectors and quantify its advantages through a cost model that tracks both mass and volume of material as a function of substrate area, film thickness, and bath geometry. In addition to lowering material costs, this strategy directly addresses the sustainability challenges posed by toxic solutions that incorporate restricted heavy metals. Coupled with our open-source, low-cost automated dip coater, the biphasic method offers a practical route for scalable and more responsible processing of cQD films.

**Materials and Methods**:

**Materials**
Lead (II) acetate trihydrate ($Pb(OAc)_2·3H_2O$, puriss. 99.5-102.0%) oleic acid (OA, 70%, technical grade), 1-octadecene (1-ODE, 90%, technical grade, bis(trimethylsilyl)sulfide (($TMS)_2S$, synthesis grade), toluene ($C_6H_5CH_3$, anhydrous, 99.8%), Hellmanex III, tetrabutylammonium iodide (TBAI, 98%), Acetone (>=99.5%) were purchased from Sigma-Aldrich. Hexamethyldisilazane (HMDS 98%) was purchased from Fischer Scientific. Methanol ($CH_3OH$, reagent grade) was purchased from Greenfield Global Inc. Sodium dodecyl sulfate (SDS, >= 99.9%) was purchased from AmericanBio.

**PbS Synthesis**
The PbS synthesis mimics the synthesis from Thompson *et al.* [24] A lead-oleate precursor was prepared by degassing 3.8 g $Pb(OAc)_2·3H_2O$, 75 mL 1-octadecene (1-ODE), and 25 mL oleic acid (OA) at 100 °C under vacuum until the mixture became optically clear. Separately, 1.05 mL bis(trimethylsilyl)sulfide (($TMS)_2S$) was dissolved in 50 mL 1-ODE inside a nitrogen glovebox. The flask was heated to 150 °C. This sulfur solution was swiftly injected into the flask. The reaction was allowed to proceed for 60 s, after which the flask was removed from the heat, air-cooled, and transferred back into the glovebox. The crude product was purified by three cycles of precipitation with methanol/acetone and redispersion in toluene, and the final PbS cQDs were stored as a toluene dispersion.

**$CsPbBr_3$ Synthesis**
$CsPbBr_3$ cQD synthesis was adapted from previous works.[25,26] In a three-necked round bottom flask, 5 mL ODE and 69 mg (0.188 mmol) of $PbBr_2$ were added and the mixture was kept under vacuum at 120 °C for 30 min followed by $N_2$ purging. This vacuum-purging step was repeated 3 times. Dried OA and OLA (0.5 mL each) were added to the mixture at 120 °C under $N_2$ flow. After the solubilization of $PbBr_2$, the reaction temperature was increased to 170 °C, and 0.8 mL of Cs-oleate (pre-heated at 80 °C) was injected into the mixture. The reaction was stopped after 5 s using an ice bath.

**Dip Coating of PbS Photoconductors**
Interdigitated Indium Tin Oxide (ITO) substrates with 50 μm spacing were obtained from South China Science & Technology Company Limited and underwent a multi-step cleaning procedure. First, they were sonicated in a boiling DI water bath containing 1% vol Hellmanex solution for 5 minutes, followed by additional sonication in neat boiling DI water and then room-temperature DI water for 5 min each. The substrates were then sonicated in acetone for 5 min. The substrates then underwent a UV-Ozone treatment for 20 min. 36 ml of a 10 mg/ml solution of PbS cQDs in toluene was prepared in a 50 ml beaker this was used as the first beaker for the 3-stage dip coating. The second beaker contained 36 ml of a 10 mg/ml solution of TBAI in methanol and the third beaker contained 36 ml of neat methanol. The dip coater was run at 1000 mm/min with a 1s hold time in

Beaker#1 followed by a 10 s hold time in Beaker#2 and a 10 s hold time in Beaker#3. This process was repeated for the specified number of dips used in the paper.

**HMDS Vapor Treatment**
Cleaned glass slides or interdigitated ITO substrates were first dehydrated on a 140 °C hot plate for 10 min to remove surface moisture. The substrates were then mounted on holders inside a vacuum desiccator above a dish containing 2–3 drops of hexamethyldisilazane (HMDS). The chamber was evacuated for 5 min, after which the pump was isolated so the slides could react with the HMDS vapor under static vacuum for 15 min. Finally, the desiccator was vented, and the slides were returned to the 140 °C hot plate for 5 min to drive off excess reagent and complete the silanization.

**Biphasic Dip Coating of PbS Photoconductors**
Interdigitated ITO substrates with 50 µm spacing were cleaned and coated with HMDS, as described above. The three beakers were then prepared as described above, with the exception that Beaker#1 was filled with 32 ml of a 6.25 µM sodium dodecyl sulfide solution in de-ionized water, followed by gentle addition of 4 ml of 10 mg/ml PbS cQDs in toluene to form the biphasic mixture. The dipcoating parameters are also identical to the monophasic case.

**Spin Coating of PbS Photoconductors**
The substrates were cleaned as described above. Then 30-50 mg/ml PbS cQDs in toluene was spincoated at 2000 rpm for 50 s followed by 10 s ligand exchange with 10 mg/ml solution of TBAI in methanol and two washes in neat methanol.

**Dip Coater Height Measurements**
The vertical position of the dip-coater arm was monitored in real time using an ultrasonic distance sensor the HC-SR04. The sensor was mounted in a fixed position identical to where the substrate would be placed in such a way that the acoustic beam was aligned with the base of the dipcoater. The trigger and echo pin of the sensor was connected to the Arduino Uno microcontroller. The Arduino generated a 10 µs trigger pulse at 1 s intervals, and the echo return time was converted to distance using the known speed of sound in air. The resulting displacement values were transmitted over USB serial communication to a host computer for logging.

**Photoresponse Measurements**
The photoresponse of the sample was characterized using a custom fabricated visible-to-SWIR photoconductivity setup. 300 - 3800 nm broadband light from an incoherent 250 W Oriel Newport Light source equipped with a halogen bulb was collimated, filtered for second order light and chopped at 25 Hz before being focused onto the input slit of a Cornerstone 260 Vis-NIR extended range ¼ m monochromator. Based on the selected monochromator wavelength, suitable high pass filters (375 nm, 715 nm or 1400 nm) were automatically selected. Maximum power output from the monochromator, while maintaining 39 nm resolution, was ensured through adjusting input and output slits to 3 mm. After collimation, the electromagnetic radiation exiting the monochromator was refocused onto the sample mounted in a dark enclosure. An 843-R-USB power meter coupled with a Germanium (Ge) reference detector (818-ST2-IR) was utilized to quantify the wavelength-dependent input power hitting the sample. The generated photocurrent was amplified and converted to a voltage output through a Stanford Research Systems SR570 current preamplifier connected in series with an SR810 lock-in amplifier, allowing for extraction of the photovoltage signal from the light -exposed sample. Sample photovoltage and phase output from the lock-in amplifier were read and saved using a custom-built LabView program.

**Fourier Transform Infrared Spectroscopy (FTIR)**
FTIR spectra are measured in the range of 11000 – 6000 cm$^{-1}$ using a Nicolet iS-50 FTIR spectrometer using the DTGS detector and white light source.

**Atomic Force Microscopy Measurements (AFM)**
A Bruker Dimension Icon Atomic Force Microscope was used in ScanAsyst Air mode to measure film roughness directly on a spin-cast film in an identical manner to the fabrication of the photoconductors. In this mode, the gain and frequency used during the measurement is optimized by the software.

**Profilometry for Thickness Measurements**
A Bruker DektakXT is used to measure film thicknesses above 30 nanometers. A small incision is made in the film using a fine tweezer, and the resulting step profile is scanned. The film thickness is extracted by averaging the baseline height on either side of the incision, excluding the local deformation produced at the cut edge.

**Scanning Electron Microscopy (SEM)**
The film morphology was characterized using a high-resolution scanning electron microscope operated at 3 kV and 100 pA (Oxford Instruments Merlin Carl Zeiss HR-SEM).

**Photoluminescence (PL)**
Emission intensities were collected using a PV Microscope Spectrophotometer from CRAIC Technologies. The sample is excited with 365 nm light. The glass rod was mounted onto the microscope using a 3D printed part and rotated to allow for radial measurements.

**Results and Discussion**:
To enable reproducible and scalable fabrication of cQD thin films, we built a dipcoater capable of automated three-stage reservoir dipping, as shown in Figures 1a–c. All components are listed in Table S1. Readily available parts were sourced from Amazon and McMaster-Carr resulting in a net cost of US $315 to fabricate the entire customized dipcoater, a cost that is substantially lower than commercial dip coaters, which typically range from US $3,000–6,000 for single- and multi-stage systems as listed in Table S2. In our system, the substrate is mounted on the dip-coating arm and programmed via the graphical user interface shown in Figure 1d, which sets the withdrawal speed to define film thickness, the dwell time to control the duration of ligand exchange and washing, and the cycle count to determine the total number of coatings. Mechanical precision is assessed by repeated height cycling measurements performed with a HC-SR04 ultrasonic sensor over an extended run of 600 cycles. Representative traces for this run are shown in Figure 1e. Across the full test, the maximum arm position drift remained below the measurement uncertainty of the height sensor of 0.3 cm, shown in Figure 1f, indicating no measurable systematic drift and confirming the long-term stability and reproducibility of the coater for multi-layer film deposition.

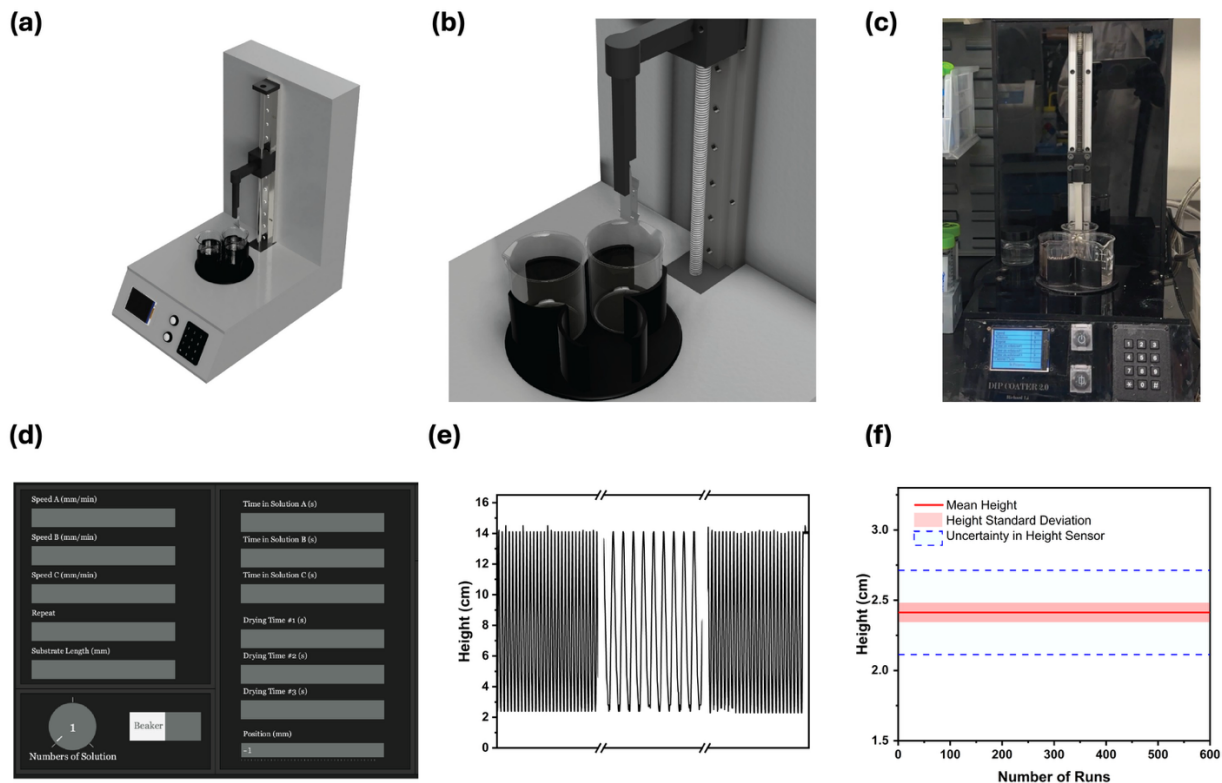

**Figure 1:** Dip coater and reliability. (a) Render of multi-stage dip coater showing the coating arm and modular rotating bath station. (b) Detailed view of the rotating bath station (c) Photograph of the constructed dip coater (d) Custom graphical user interface enabling independent control of speed, dwell time and stage position for dip coater (e) Recorded height profile of dip coater arm for 600 consecutive runs, demonstrating repeatable motion with no drift. (f) Bottom-position stability over 600 cycles, showing that any variations remain within the intrinsic uncertainty of the height sensor.

To test the efficacy of our dipcoater for cQD thin film deposition, we implemented an automated analogue of the standard layer-by-layer solid-stage ligand exchange used to enhance electronic coupling in cQD films.[27] The process is depicted in Figure 2a, where each cycle begins with the deposition of long-ligand-capped cQDs onto the substrate during its withdrawal from the cQD dispersion (Stage 1). The dried film then undergoes a ligand exchange in a short-chain ligand bath (Stage 2), which displaces the native electrically insulating ligands, compacts the film and strengthens interparticle coupling.[4,27] A subsequent rinse in a neat non-solvent (Stage 3) removes the displaced ligand that would impede charge transport. This sequence is identical to the conventional spin-coating workflow, where the spin, exchange and wash steps are executed manually in series.[28] However, building a 500 nm cQD film for an optoelectronic device by spin-coating typically requires over a dozen such layer-by-layer cycles, making the process time consuming and operator intensive. By contrast, the dip coater executes the entire sequence autonomously and can accommodate multiple substrates on the coating arm, as shown in Figure S2, substantially reducing hands-on time and improving process reproducibility.

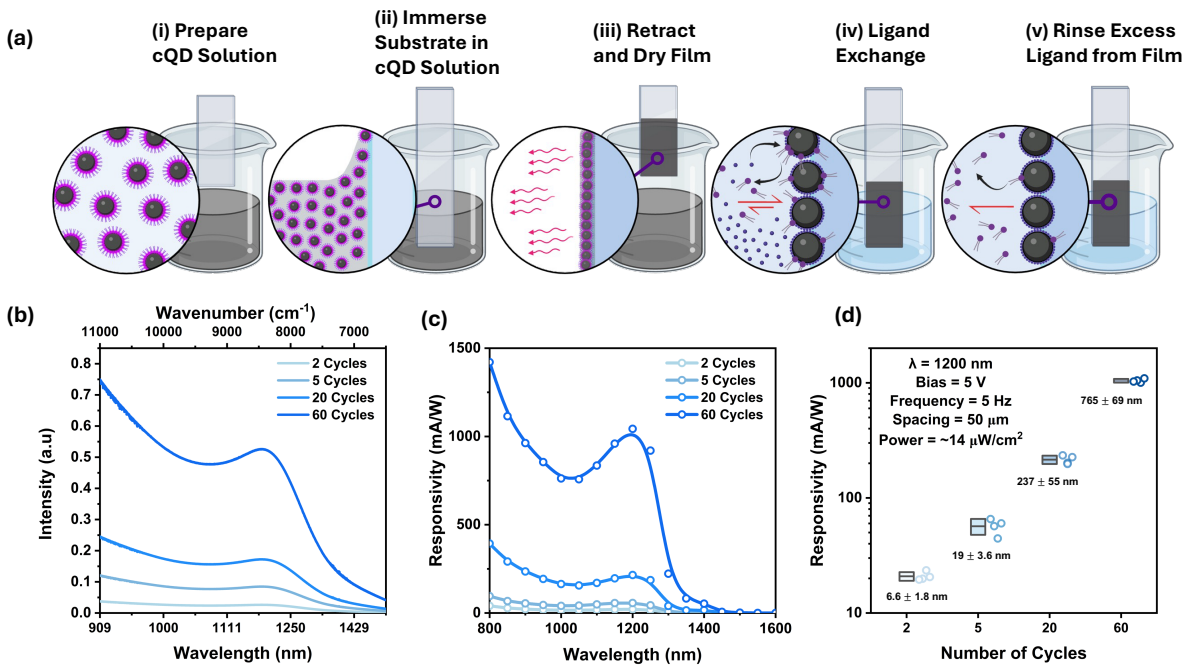

**Figure 2:** (a) Schematic of the automated dip-coating workflow, illustrating sequential nanoparticle deposition, film drying, ligand exchange and removal of excess ligands (b) Absorption spectra of PbS cQD films showing monotonic increase in optical density as the number of dip coating cycles increases, consistent with film thickening. (c) Spectral responsivity of PbS photodetectors fabricated with varying cycle counts, demonstrating enhanced responsivity with progressive film growth. (d) Statistically reproducible responsivity values at 1200 nm as a function of number of dip coating cycles. Thickness values and standard deviations for various dipping cycles are presented below each box plot.

To benchmark the functional performance of dip coated films, we fabricated photoconductors using PbS cQD films of varying thickness sandwiched between ITO electrodes. In Figure 2b we show that increasing the number of dips results in an increase in the near infrared excitonic absorption of the PbS cQD films, confirming both the film buildup and preservation of the quantum dot confinement post cQD deposition and subsequent ligand exchange. Spectral responsivity measurements shown in Figure 2c reveal broadband photoresponse extending from 800 nm to 1400 nm. The photoresponse increases monotonically with increasing number of dips. We attribute this to an increase in film thickness, consistent with enhanced absorption observed in Figure 2b. We repeat the same experiment for multiple samples to test the reproducibility of the approach. Figure 2d depicts how the responsivity tracks across various samples at various number of dips. The mean responsivity values increase from ~20 mA/W for 2-cycle devices to 1000 mA/W for 60-cycle devices at 1200 nm under 5V applied bias and a 5 Hz chopping frequency. We will compare the thickness normalized responsivity against other coating methods in a later section of the text.

While these results establish that dip-coated films provide reproducible and thickness-tunable optoelectronic performance, the broader challenge lies in scaling such processes for large-area fabrication. Conventional single-phase or monophasic dip-coating requires an entire reservoir of cQD solution deep enough to submerge the substrate, yet only the thin layer at the liquid–air interface contributes to film formation.[15] As a result, most of the active solution, which can be costly or toxic, in a monophasic bath remains unused. To overcome this inefficiency, we employ a biphasic dipcoating strategy in which a denser, immiscible phase supports a thin upper phase of active solution.[14] During immersion and withdrawal, the substrate passes through both phases, yet film formation occurs exclusively at the liquid–air interface of the active phase.[14] This method dramatically reduces the volume of cQD solution required, while maintaining film thickness. Choosing an appropriate sacrificial solvent requires balancing several attributes: it must exceed the active cQD solvent in density, remain immiscible with the active phase, exhibit chemical inertness, and resist evaporation during processing. [14]

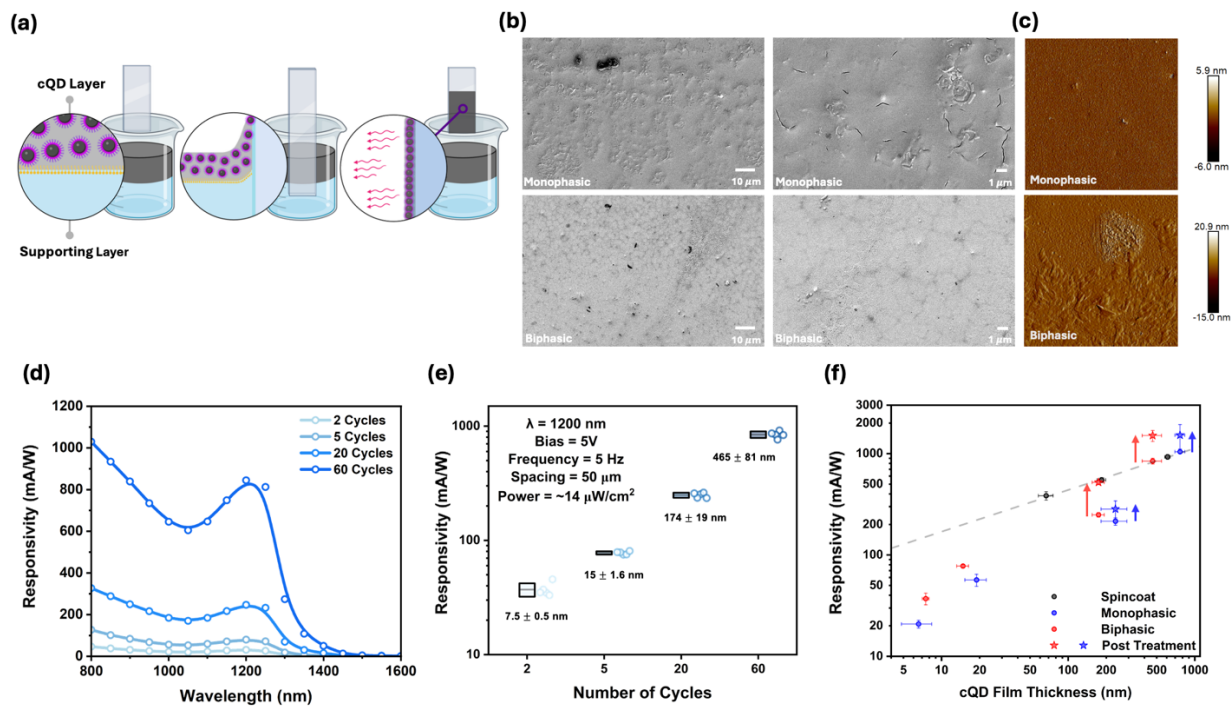

**Figure 3**: (a) Schematic of the biphasic dip-coating process, in which the cQD solution is supported by a denser immiscible underlayer, DI water in this case, enabling metered withdrawal of a thin film of PbS cQDs. (b) Scanning electron microscopy images of ~200 nm thick 20-cycle monophasic and biphasic films showing contraction-induced cracking characteristic of the solid-state ligand exchange process. (c) Atomic force microscopy height maps comparing film topography for monophasic and biphasic depositions with corresponding average roughness values of 1.74 nm and 4.59 nm respectively. (d) Spectral responsivity curves of biphasic dipcoated PbS photodetectors with increasing dip cycle count. (e) Statistical representation of responsivity values at 1200 nm as a function of biphasic dip-coating cycle number. Thickness values and standard deviations for thickness are presented below each box plot demonstrating good reproducibility. (f) Responsivity at 1200 nm and 5V as a function of film thickness for spin-coated, monophasic dip-coated and biphasic dip-coated devices. The grey dashed line denotes the linear trend of traditional spin-coated devices, while arrows indicate the performance gains achieved after an additional 20-second ligand exchange on some of the dipcoated films indicating that performance can match standard spin-coated devices when further optimization is performed.

Figure 3a outlines this approach, wherein a thin nonpolar layer of PbS cQD dispersion is metered atop a sacrificial solvent underlayer. Prior work by Grosso *et al.* used dense and inert perfluorohexanes (PFH) as an underlayer to coat $SnO_2$, $TiO_2$ and polymer films.[14] However, perfluoroalkyl substances such as PFH are environmentally persistent, bioaccumulative and increasingly regulated, complicating waste handling and sustainability goals.[29] Here we use deionized water, which is immiscible with the toluene-based cQD solution, with 6.25 µM sodium dodecyl sulfate added to lower surface tension at the water-toluene interface.[30] This suppresses entrainment of water onto the substrate during withdrawal which can lead to extremely patchy films.

Representative surface morphology of the 20-cycle monophasic and biphasic films is shown in Figure 3b and additional SEM images of the 20- and 60- cycle films are provided in Figures S3-S10. The films show characteristic cracks from film densification during solid-state ligand exchange, as reported previously for similar systems.[31,32] The roughly 500 nm thick 60-cycle biphasic films exhibit spherical defects about 2–3 µm in size attributed to localized water entrapment during film formation as seen in Figures S5-S6.[33] Atomic force microscopy analysis over a 25 µm² area, shown in Figure 3c, yields average surfaces roughness values of ~1.74 nm for monophasic and ~4.59 nm for biphasic 20 cycle films, where the increased roughness is consistent with the increased interfacial complexity introduced by the two-phase system. The responsivity spectra, shown in Figure 3d of the biphasic devices closely resembles that of the monophasic devices shown in Figure 2, confirming that the biphasic approach preserves the intrinsic photoconductivity of the cQD film. Statistical device data, shown in Figure 3e, demonstrate good reproducibility across multiple fabrications.

To rationally compare device metrics across thicknesses and deposition methods, we normalize performance to the active film thickness rather than the number of dips. We therefore measured the thickness of each film at five locations across the substrate and fabricated a set of spin-coated devices using the established layer-by-layer protocol.[34,35] The monophasic and biphasic devices achieve comparable responsivities confirming that the material efficiency of the method does not compromise its performance. At very low thicknesses, we observe larger variations in device performance, which we attribute to incomplete substrate coverage. Atomic force microscopy reveals island-like morphologies for films thinner than 10 nm. With increasing number of dips, these voids are subsequently filled and a uniform continuous network of cQDs is formed, leading to improved charge transport with less variability in responsivity. Interestingly, our spin coated films show better performance than our dipcoated films at comparable thicknesses. We hypothesized that this discrepancy might be due to unoptimized ligand exchange in our dipcoated films. We note that spin-coating protocols for PbS cQD devices have benefited from nearly fifteen years of optimization.[36,37] To test this hypothesis, we soaked the fabricated dip-coated devices in the ligand-exchange bath for an additional 20 s. Invariably across all samples, we observe an increase in the spectral responsivity with some values surpassing that of spin-coated devices as shown by the post-treatment data (stars) in Figure 3f. These results establish that, with modest optimization, both monophasic and biphasic dip coating can serve as viable cQD film fabrication approaches, yielding device performance comparable to spin coating while consuming only a fraction of the active material.

To quantify the material and economic implications of each deposition route, we constructed a mass–volume–cost model parameterized by substrate diameter, target film thickness and production scale. The underlying expressions and assumptions are detailed in Section S3. In brief, the model first determines the ideal dry film mass from the substrate diameter, the effective nanocrystal density and the desired thickness. Process losses are then incorporated by assigning conservative material utilization efficiencies of 5% for spin coating and 90% for dip coating. For dip-based methods, the total volume of active solution additionally includes the reservoir required to fill the coating bath, which scales with the footprint of the substrate. In the biphasic configuration only the cQD rich upper layer is counted as active, whereas the immiscible underlayer is treated as an inexpensive supporting phase. The analysis focuses on cQD dispersion consumption and cost and is therefore intended to isolate the regimes in which the material efficiency of dip coating compensates for the reservoir overhead.

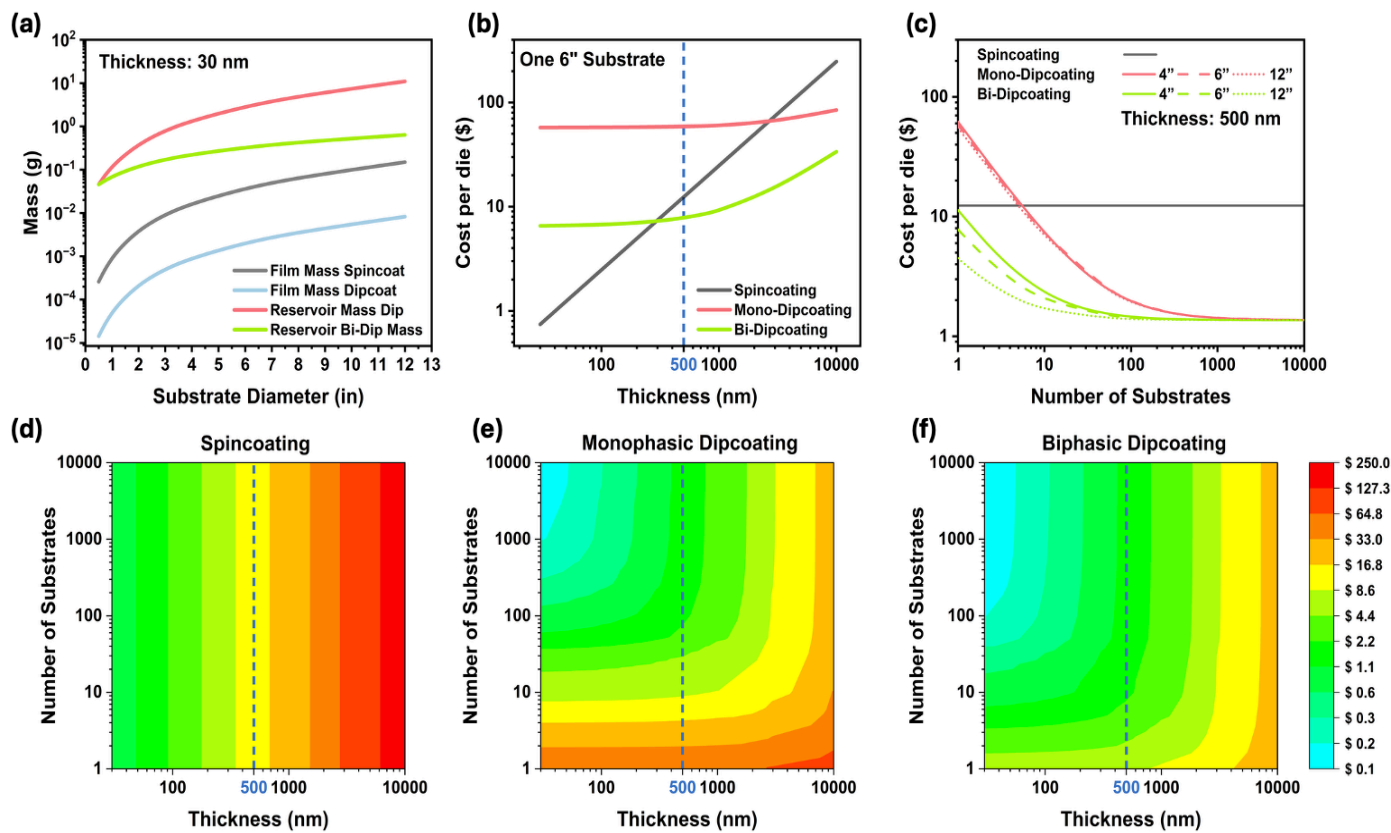

**Figure 4:** (a) Calculated film and reservoir mass requirements as a function of substrate diameter for a 30nm coating. (b) Calculated cost per die for a 6" substrate as a function of coating thickness. (c) Cost per die as a function of batch size for 4" 6" and 12" substrates at a fixed thickness of 420 nm. The benefit of dipcoating becomes pronounced at larger scales due to retention into the bath of excess material. (d) Parametric cost-surface maps showing cost per die for spincoating (e) Monophasic dipcoating (f) Biphasic dipcoating. The biphasic method maintains the lowest overall cost and exhibits a broad low-cost plateau highlighting its material efficiency.

We first examine, in Figure 4a, the total active mass consumed for a representative 30 nm film. When the reservoir contribution is neglected, spin coating requires approximately one order of magnitude more cQD mass than dip coating to reach the same thickness, as expected from its low utilization efficiency. For a 6-inch substrate, the required mass is 37 mg for spin coating compared with 2 mg for dip coating. When the reservoir is included for dip coating, the initial cQD mass required to perform the deposition becomes strongly thickness and geometry dependent. For thin films on a single wafer, the active ink needed to fill the bath dominates the material budget and can exceed the mass of cQDs required to fabricate the film by several orders of magnitude. This illustrates why dip coating is often perceived as inefficient when considered in isolation for thin films. The plots for the volume scaling, showing a similar trend as the mass, are provided in Figure S12.

We next consider how this comparison evolves with film thickness. Figure 4b reports the cost per die as a function of thickness for a single 6-inch substrate. Here a die is considered to be 14.6 mm (~ 6-inch) wide and long.[38] Optimal thicknesses for the active cQD layer in PbS photodiodes are reported to be between 450 – 600 nm, to illustrate this optimum in the context of the model, we add a dashed blue line representing 500 nm.[39–41] For spincoating, the cost increases linearly with thickness since each additional cycle requires fresh solution. For the dip-based methods, the cost decomposes into a thickness-independent term associated with initial filling of the active bath and a thickness-dependent term associated with replenishing the bath as more material is withdrawn. At small thicknesses, the reservoir term dominates and both monophasic and biphasic curves appear nearly flat. Their vertical separation reflects the difference in active bath volume since the biphasic configuration requires an order of magnitude less cQD dispersion in its upper phase than the monophasic bath and therefore starts from a much lower cost per die. Owing to its smaller active reservoir, the biphasic method shifts the break-even thickness for dipcoating by an order of magnitude within the technologically relevant thickness window of 450-600 nm.

Throughput introduces a further dimension. Figure 4c shows the cost per die as a function of the number of substrates processed, at a fixed thickness of 500 nm and for wafer diameters of 4-, 6- and 12- inches. Once the dip coating reservoir is in place, repeated use of the same bath causes the cost per die to fall rapidly with increasing batch size and then saturate at a limiting value where only incremental replenishment of the bath is significant. Spin coating lacks this amortization since each substrate requires an unrecoverable batch of solution and so the cost per die remains nearly independent of batch size. The biphasic traces approach their steady state value at substantially smaller batch sizes than the monophasic traces, consistent with the reduced active bath volume that must be recovered.

Finally, Figures 4d–f present cost surface maps as a function of thickness and batch size for spin coating, monophasic and biphasic dip coating on 6-inch wafers. The spin coating surface shows that cost scales strongly with thickness and exhibits no relief with increasing throughput, reflecting the intrinsic scale invariance of a single use process. Monophasic dip coating introduces a valley of reduced cost at large thickness and large batch size, but the low-cost region is constrained by the need to maintain a large active bath. In contrast, the biphasic surface displays a broad and deep low-cost basin extending to both thinner films and smaller batch sizes. In this regime the cost per die is largely decoupled from device scale or geometry and is instead governed by the small volume of active solution required to replenish the upper phase. Taken together, these calculations show that the biphasic geometry shifts the locus of economic viability of dip coating toward the thicknesses and throughputs that are characteristic of practical cQD devices.

We next examine whether the biphasic approach can be extended beyond PbS cQDs to other cQDs and nanocrystals and to non-planar substrates where spin coating cannot be employed. Figure 5a shows a representative bath in which $CsPbBr_3$ nanocrystals in toluene form the active upper phase above an immiscible perfluorohexane underlayer. For $CsPbBr_3$, water cannot serve as the sacrificial phase because its high polarity strips weakly bound ligands, promotes hydration and ion solvation, and rapidly decomposes the perovskite lattice.[42–44] Existing solution based methods such as spray coating and inkjet printing have begun to relax the constraint of planar substrates for perovskite and colloidal nanocrystal devices.[45–48] Dip coating provides a complementary route in which film formation is governed by capillary entrainment rather than droplet flux, enabling conformal coverage over diverse length scales while retaining the high material efficiency established above. [13,49] As a proof of concept for non-planar biphasic coating, we immersed an HMDS treated glass rod into the CsPbBr$_3$/perfluorohexane bath and withdrew it at 400 mm/min as illustrated in Figure 5b. This process produced a continuous emissive film on the cylindrical surface as shown in Figure 5c. To assess circumferential uniformity, the rod was mounted in a custom photoluminescence microscope, and the emission intensity was recorded as a function of the azimuthal angle. The resulting polar plot in Figure 5d reveals less than ten percent variation in photoluminescence intensity around the cylinder, providing an initial proof of uniform coating of solution-processed nanocrystals on curved surfaces. Pushing to higher curvatures, HMDS treated glass marbles were coated under analogous conditions and exhibit bright emission under 405 nm illumination although quantitative mapping on the sphere was not possible due to difficulties in focusing the microscope on the extreme curvature. We believe that these initial investigations might enable the development of exotic and relatively unexplored optoelectronic devices such as curved light emitting diodes, domed photodetectors and curvature conforming solar cells.[50–52]

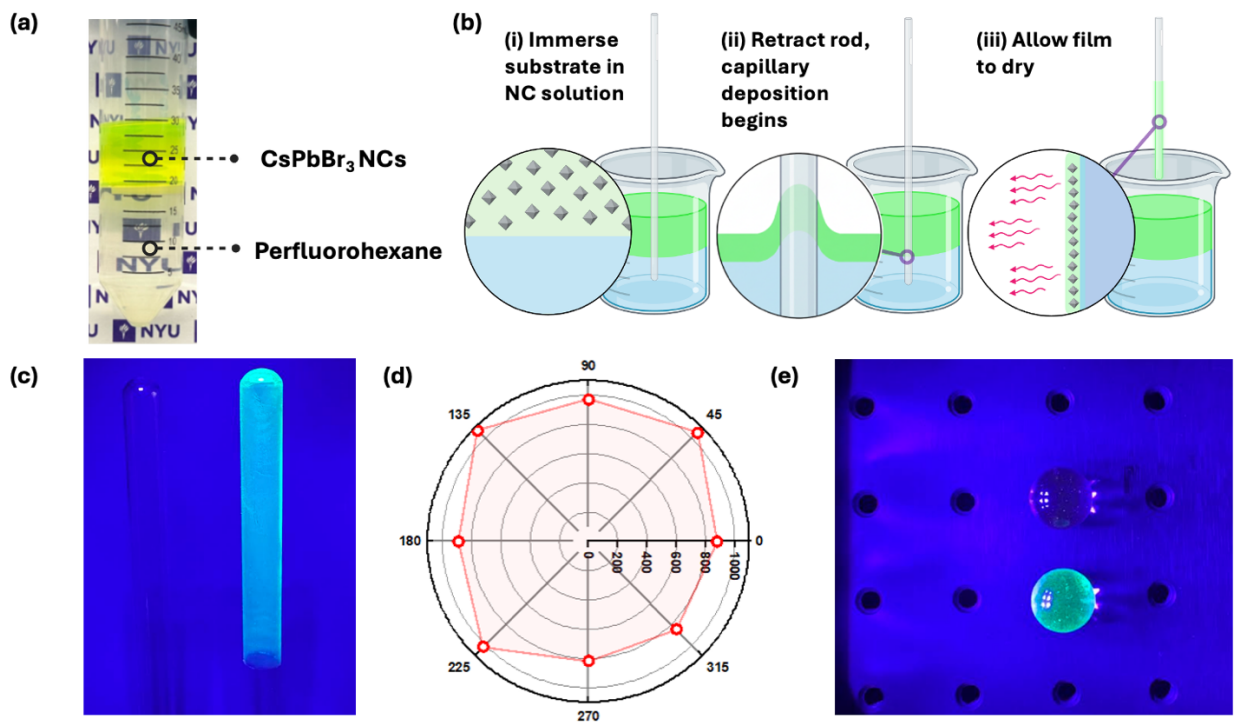

**Figure 5:** (a) Photograph of CsPbBr$_3$ NCs suspended on top of perfluorohexanes demonstrating feasibility of the biphasic method to coat perovskite NCs. (b) Schematic of the coating process for non-planar substrates, here a glass rod is fully immersed into the biphasic bath and withdrawn at a controlled rate to form a nanocrystal layer coating. (c) UV-illuminated image of glass rods with and without CsPbBr$_3$ NC coating. (d) Polar plot of photoluminescence intensity as a function of azimuthal angle around the rod, confirming near-isotropic emission highlighting the potential for uniform circumferential coverage. (e) UV-illuminated image of coated and uncoated glass marbles, demonstrating the adaptability of the biphasic method for extreme curvature deposition.

**Conclusions**:
We have introduced an automated dip-coating platform that delivers reproducible, device-grade photo-active cQD films with minimal material use. By implementing a biphasic reservoir that replaces most of the active ink with an immiscible underlayer, we reduce solution consumption by nearly an order of magnitude while maintaining film quality and photodetector performance. A mass–volume analysis shows that this geometry shifts the cost minimum of dip coating into the thickness and throughput range relevant for modern optoelectronic devices. The same workflow extends to perovskite nanocrystals and to curved substrates, highlighting its versatility. Together these results position biphasic meniscus coating as a practical and material efficient alternative to spin coating for scalable solution-processed optoelectronics.

**Author Contributions**:
S.J.P.: Investigation, conceptualization, methodology, formal analysis, visualization, resources, data curation, writing – original draft, writing – review and editing. L.L.: resources, visualization, data acquisition. Z.L, A.K and M.K.: data acquisition. S.S.L.: writing – review and editing. A.S.: conceptualization, project administration, funding acquisition, supervision, methodology, validation, writing – review and editing.

**Acknowledgements**:
This study was funded by the Young Faculty Award program of the Defense Advanced Research Projects Agency (DARPA) under the grant D21AP10118. The views, opinions, and/or findings expressed are those of the authors and should not be interpreted as representing the official views or policies of the Department of Defense or the U.S. Government. This work was also supported by the Office of Naval Research grants N00014-20-1-2231 and N00014-24-1-2683. We also gratefully acknowledge support for instrument use, scientific and technical assistance from the NYU Shared Instrumentation Facility through the Materials Research Science and Engineering Center (MRSEC) and MRI programs of the National Science Foundation under Award numbers DMR-1420073 and DMR-0923251. The authors acknowledge the NYU Tandon Makerspace for fabrication support. This work was performed [in part] at the NYU Nanofabrication Cleanroom Facility (NYU Nanofab).

**Notes:**
The authors declare the following competing financial interest(s): A.S., L.L. and S.J.P. have filed a provisional patent application based on the results of this manuscript.

**Data Availability Statement:**
The data that support the findings of this study are available from the corresponding author upon reasonable request.

**Supporting Information:**
Supporting Information is available below.

**TOC Image:**

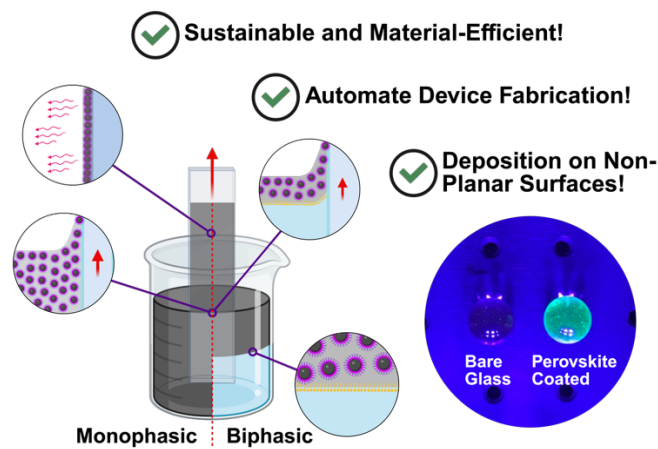

**TOC Text:**
This study introduces an automated dip-coating method that produces uniform colloidal quantum dot films while markedly reducing material consumption. By supporting a thin active solution on an immiscible underlayer, the process lowers solution use by nearly an order of magnitude without compromising film quality. The approach accommodates diverse nanomaterials and curved substrates, offering a practical route toward scalable, low-waste optoelectronic fabrication.

# Supporting Information

**Title**: Biphasic Meniscus Coating for Scalable and Material Efficient Nanocrystal Films

**Author List**: Shlok Joseph Paul,[1] Letian Li,[1] Zheng Li,[1] Andrew Kim,[1] Mia Klopfestein,[2] Stephanie S. Lee,[2] Ayaskanta Sahu*[1]

**Affiliations**:
[1] Department of Chemical and Biomolecular Engineering, NYU Tandon School of Engineering, 6 Metrotech Center, Brooklyn, NY, 11201, USA
[2] Molecular Design Institute, Department of Chemistry, New York University, New York, NY, 10012, USA

* Please forward all correspondence to asahu@nyu.edu

## Section S1: Building the dipcoater

All information regarding building the dipcoater is available at our GitHub: https://github.com/Richard663168/Dip-Coater

**Table S1**: Build of Materials for dipcoater

| Item | Quantity | Reference Price | Source |
|---|---|---|---|
| Arduino Mega 2560 | 1 | $20.99 for 1 | Amazon |
| ST7789 2.4" SPI LCD | 1 | $6.99 for 1 | Taobao |
| Nema 17 Stepper Motor | 1 | $10.99 for 1 | Amazon |
| 2.1x5.5 12V Power Barrel | 1 | $1.24 for 1 | McMaster-Carr |
| DM542T Stepper Motor Driver | 1 | $28.99 for 1 | Amazon |
| DM320T Stepper Motor Driver | 1 | $15.99 for 1 | Amazon |
| 3x4 Matrix Keypad | 1 | $6.50 for 1 | Adafruit Industries |
| Screw Terminal Block Breakout Module for Arduino Mega | 1 | $32.00 for 1 | Amazon |
| FUYU FSL30 Mini Linear Stage Small Slide | 1 | $118.00 for 1 | Amazon |
| 8-channel Bi-directional Logic Level Converter | 1 | $7.95 for 1 | Adafruit Industries |
| 12V Push Button | 1 | $15.99 for 10 | Amazon |
| 1 inch OD Steel Disk | 1 | $18.99 for 50 | Amazon |
| 1/2 inch OD Steel Disk | 1 | $16.99 for 80 | Amazon |
| 1 inch OD Neodymium magnet | 1 | $8.57 for 1 | McMaster-Carr |
| 1/2 inch OD Neodymium magnet | 1 | $4.89 for 1 | McMaster-Carr |

**Table S2**: Cost of other available dipcoater instruments

| Brand | Cost | Link |
|---|---|---|
| Ossila | $3000 | https://www.ossila.com/products/dip-coater |
| Holmarc HO-TH-01 | $3010 | https://www.holmarc.com/dip_coating_unit.php |
| Holmarc HO-TH-02MD (6 stage coater) | $5559 | https://www.holmarc.com/multiple_dip_coater.php |
| MTI Corp PTL-MM02 | $4598 | https://mtixtl.com/products/millimeter-grade-programmable-dip-coater-with-touch-screen-digital-controller-1-200-mm-min-ptl-mm02 |

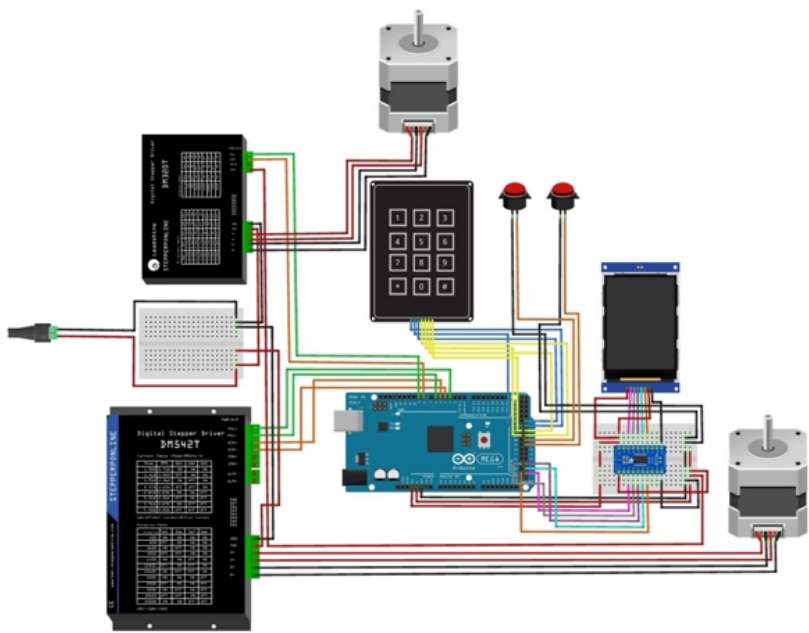

**Figure S3**: Wiring diagram for dipcoater.

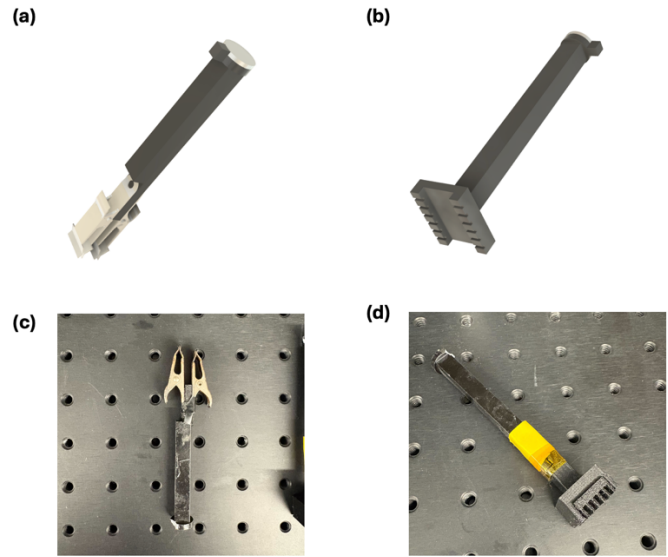

**Figure S4**: Multi-substrate attachments for dipcoater (a) Two substrate clip attachment render. (b) Multiple substrate flexible TPU holder. (c) Physical print of two substrate clip. (d) Physical print of multiple substrate flexible TPU holder. All files are available on the GitHub repository for the dipcoater.

Section S2: Monophasic and Biphasic device SEM images

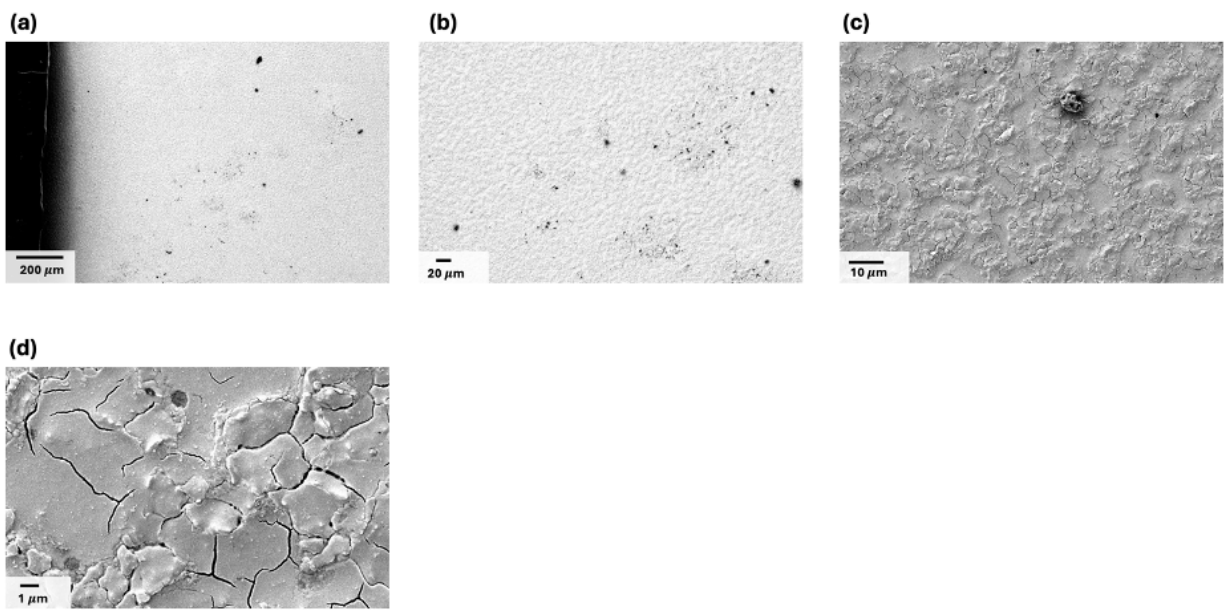

**Figure S5**: 60 dip monophasic PbS films at various magnifications: Set 1

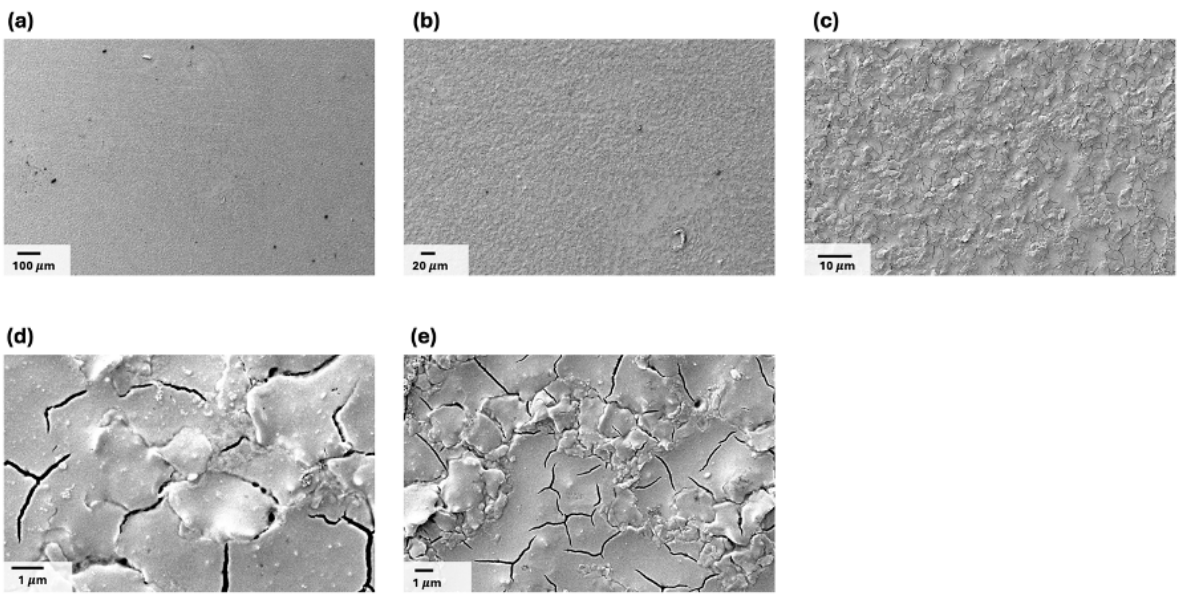

**Figure S6**: 60 dip monophasic PbS films at various magnifications: Set 2

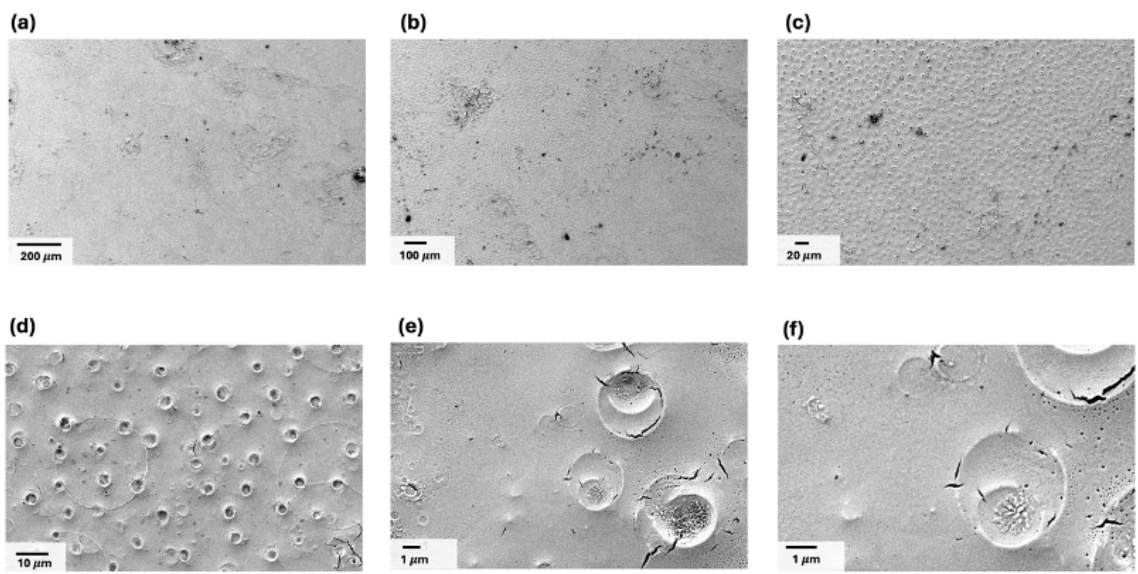

**Figure S7**: 60 dip biphasic PbS films at various magnifications: Set 1

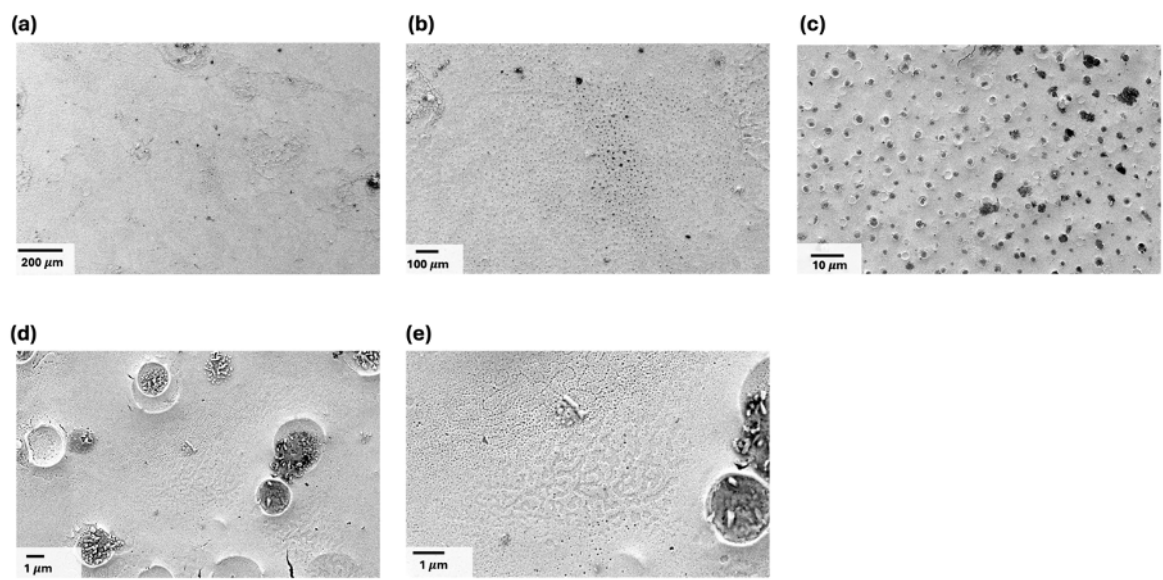

**Figure S8**: 60 dip biphasic PbS films at various magnifications: Set 2

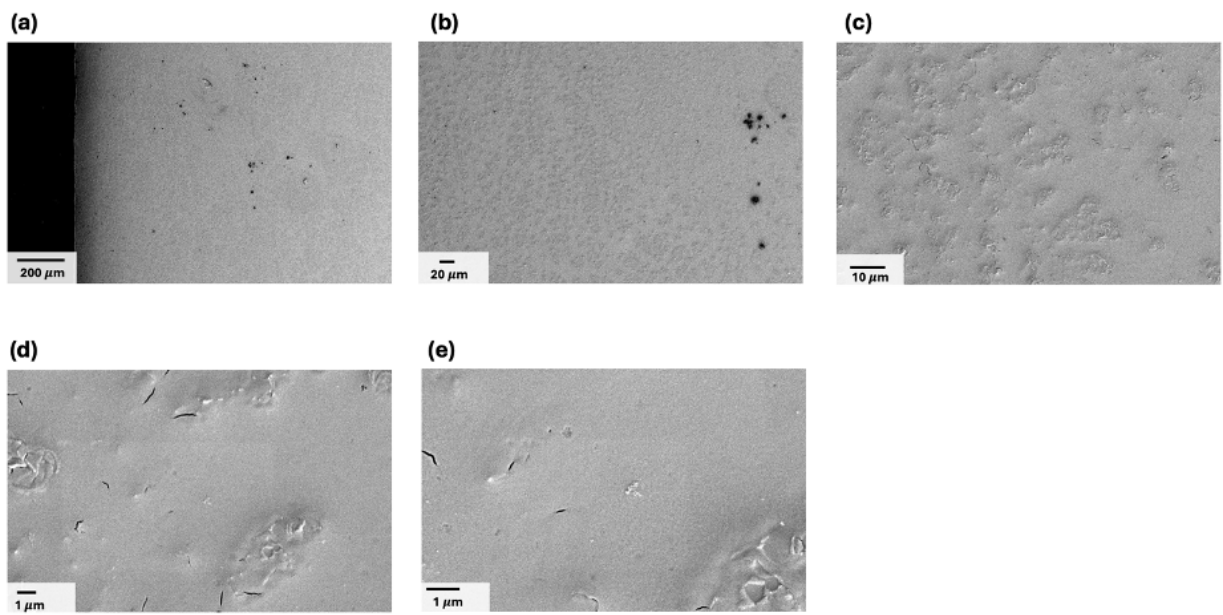

**Figure S9**: 20 dip monophasic PbS films at various magnifications: Set 1

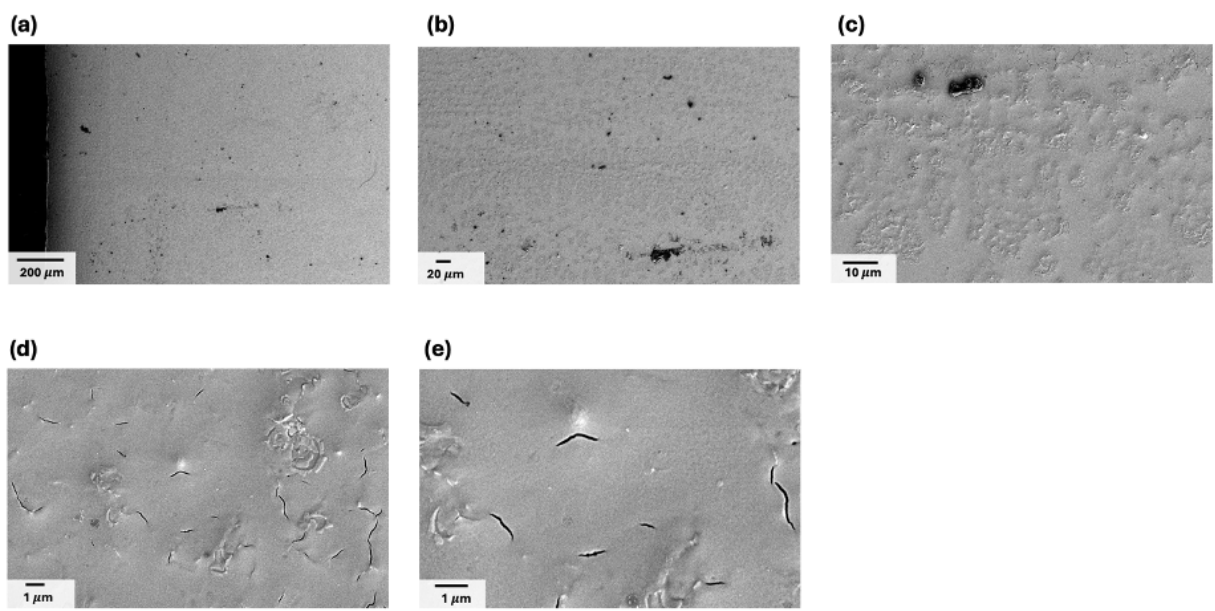

**Figure S10**: 20 dip monophasic PbS films at various magnifications: Set 2

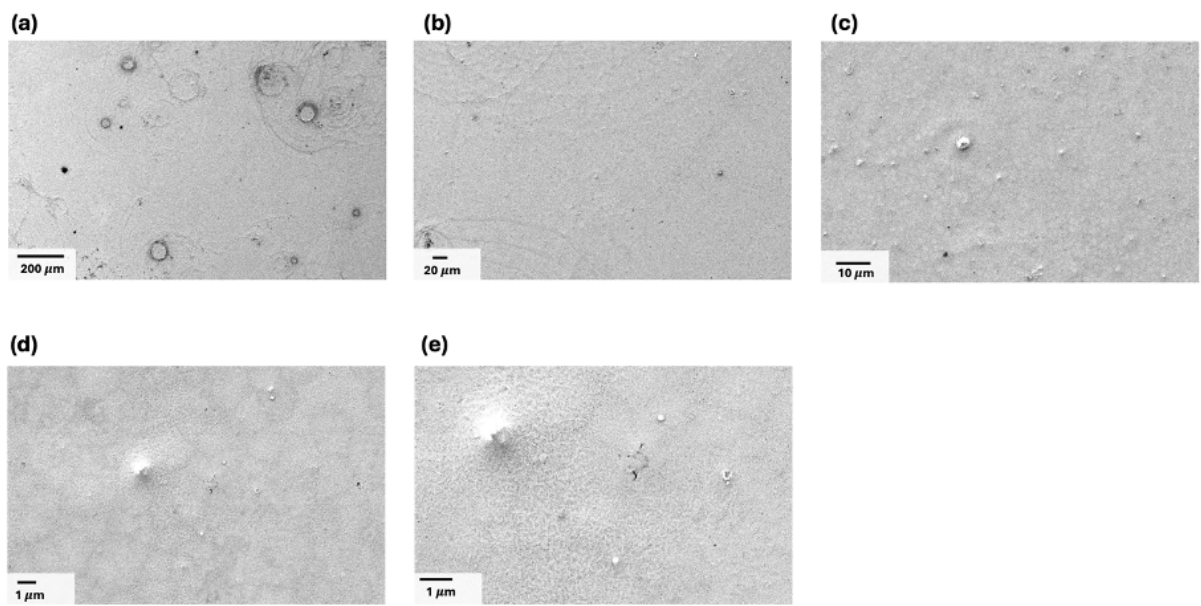

**Figure S11**: 20 dip biphasic PbS films at various magnifications: Set 1

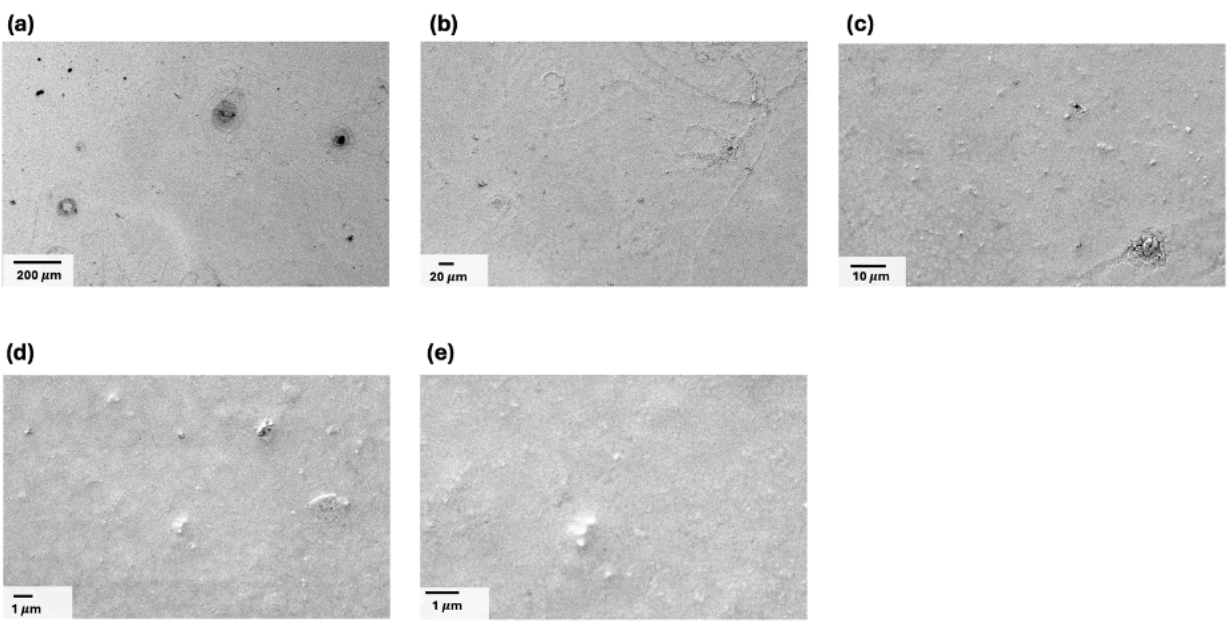

**Figure S12**: 20 dip biphasic PbS films at various magnifications: Set 2

**Section S3: Technoeconomic Analysis for coating methods**

In this section we cover the technoeconomic analysis in more detail.

All parameters used are well documented in the attached cost analysis spreadsheet. Users can tune the input parameters depending on their use cases.

To begin with we start with a mass balance.

$$M_{film} = \rho_{film} \times A \times h$$

Here A is the substrate area (cm²), h is the final film thickness (cm) and $\rho_{film}$ is the effective density of the dry film. To obtain the dry film density we start with the bulk density of PbS at 7.6 g/ml and a particle diameter (PD) of 5.2 nm typical of 1200 nm absorbing PbS cQDs. Using a packing fraction (PF) of 0.74 and a particle spacing (PS) of 1.5nm for oleic acid capped PbS, similarly used in papers modelling the refractive index of PbS we can calculate the core volume fraction.[1,2]

The core volume fraction (CVF) is given by:

$$PF \times (PS/(PD + PS))^3 = 0.74 \times (1.5/(5.2 + 1.5))^3 = 0.346$$

Now, we can isolate the contribution to density from the core PbS and the OA ligands. Using the ligand density of 1.2 g/ml for oleic acid we can compute the film density from the following equation:

$$\text{Effective Film Density} = \left(CVF \times \rho_{film}\right) + \left((1 - CVF) \times \rho_{ligand}\right) = \sim 3.414 \ g/ml$$

To calculate the mass actually used, we need to account for the inefficiencies of spincoating and dipcoating. Using conservative estimates, we assume that dipcoating has a 10% waste penalty while spincoating has a 95% waste penalty. To obtain the actual mass used to fabricate the film now we can modify the mass equation above:

$$M_{film} = [\rho_{film} \times A \times h]/\eta_{process}$$

where $\eta_{process}$ is the efficiency of the process so 5% for spincoating and 90% for dipcoating.

Now that we have the mass of the film, we also need to account for the mass required to perform the dipcoating procedure. To do this we assume that the bath is a container with dimensions like the substrate but with a margin (m%), a dead height (H) and a slot width (b) to account for the substrate needing space within the container. For the biphasic case, we denote $H_{Active}$ as the active height for the active material, since it floats on top of the supporting solvent. There is no need for the dead height term in the biphasic method since the bottom phase is insignificant. We then proceed to calculate the volume of both baths:

$$Volume_{Monophasic-bath} = (d + md) \times b \times (d + H)$$

$$Volume_{Biphasic-bath} = (d + md) \times b \times H_{Active}$$

We can calculate the mass required to populate the bath with the concentration of the solvent and the volume of the bath.

This comes out to be:

$$M_{Bath/Active} = c \times Volume$$

With both the mass of the film and mass of the bath we can calculate the total monophasic and biphasic dip mass consumption as:

$$M_{Total} = M_{film} + M_{bath}$$

With the mass balance complete we begin the solution volume balance. Spincoating typically uses a 50 mg/ml solution while dipcoating uses a 10 mg/ml solution. We start with calculating the solids volume fraction ($\phi$):

$$\phi = c/\rho_{effective}$$

Here, c is the concentration and $\rho_{effective}$ is the effective density calculated above. Now we can calculate the theoretical volume, volume for dipcoating and volume for spincoating that would be deposited on the substrate using:

$$V_{film} = (A \times T)/\phi$$
$$V_{spin} = V_{film}/\eta_{spin-process}$$
$$V_{dip} = V_{film}/\eta_{dip-process}$$

With the mass and volume balances we can calculate the cost to fabricate devices on substrates of various geometries, thicknesses and production scales. For the cost of PbS we use the cost available from Quantum-Solutions of $1700/g and for the cost of toluene we use ~ $21/L. The costs are calculated as follows:

$$Cost_{film,PbS} = M_{film} \times Cost_{PbS}$$
$$Cost_{film,Tol} = V_{film} \times Cost_{Tol}$$
$$Cost_{Bath,PbS} = M_{Bath} \times Cost_{PbS}$$
$$Cost_{Bath,Tol} = V_{Bath} \times Cost_{Tol}$$

Thus, the total cost can be represented as the sum of these costs. Since the dipcoated substrate will reduce the volume and mass of the bath with each subsequent substrate, we add back the cost to replenish the bath for solvent and mass lost. We assume that a typical die is 14.6 mm x 14.6 mm as highlighted in the technoeconomic analysis by Greboval *et al.* We can then calculate the number of dies achieved based on the substrate size being coated in order to calculate the normalized cost per die as reported in the main text.[3]

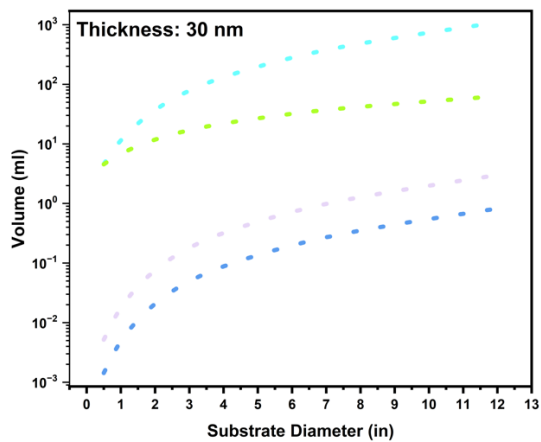

**Figure S13**: Calculated film and bath volume requirements as a function of substrate diameter for a 30nm coating